\documentclass[aps,pra,twocolumn, notitlepage,10pt,superscriptaddress,longbibliography]{revtex4-1}
\usepackage{graphicx}
\usepackage{times,bbm,amsmath,amssymb}
\usepackage{hyperref}
\usepackage{color}
\usepackage{xcolor}
\usepackage{graphicx}
\usepackage{dcolumn}%
\usepackage{bm}
\usepackage{soul}

\usepackage{nicefrac, xfrac}
\usepackage{amssymb}
\usepackage{pifont}

\def\GM#1{{\textcolor{green}{\textbf{G:}\em #1}}}
\def\CV#1{{\textcolor{red}{\textbf{V:}\em #1}}}

\usepackage{physics}

\begin{document}

\title{Quantum Optical Neuron for Image Classification via Multiphoton Interference}

\author{Giorgio Minati}
\affiliation{Dipartimento di Fisica, Sapienza Universit\`{a} di Roma, Piazzale Aldo Moro 5, I-00185 Roma, Italy}

\author{Simone Roncallo}
\affiliation{Dipartimento di Fisica, Universit\`{a} degli Studi di Pavia, Via Agostino Bassi 6, I-27100 Pavia, Italy}

\author{Simone Scrofana}
\affiliation{Dipartimento di Fisica, Sapienza Universit\`{a} di Roma, Piazzale Aldo Moro 5, I-00185 Roma, Italy}

\author{Angela Rosy Morgillo}
\affiliation{Dipartimento di Fisica, Universit\`{a} degli Studi di Pavia, Via Agostino Bassi 6, I-27100 Pavia, Italy}

\author{Nicol\'{o} Spagnolo}
\affiliation{Dipartimento di Fisica, Sapienza Universit\`{a} di Roma, Piazzale Aldo Moro 5, I-00185 Roma, Italy}

\author{Chiara Macchiavello}
\affiliation{Dipartimento di Fisica, Universit\`{a} degli Studi di Pavia, Via Agostino Bassi 6, I-27100 Pavia, Italy}

\author{Lorenzo Maccone}
\affiliation{Dipartimento di Fisica, Universit\`{a} degli Studi di Pavia, Via Agostino Bassi 6, I-27100 Pavia, Italy}

\author{Valeria Cimini}
\email{valeria.cimini@uniroma1.it}
\affiliation{Dipartimento di Fisica, Sapienza Universit\`{a} di Roma, Piazzale Aldo Moro 5, I-00185 Roma, Italy}

\author{Fabio Sciarrino}
\affiliation{Dipartimento di Fisica, Sapienza Universit\`{a} di Roma, Piazzale Aldo Moro 5, I-00185 Roma, Italy}

\begin{abstract}

The rapid growth of machine learning is increasingly constrained by the energy and bandwidth limits of classical hardware. Optical and quantum technologies offer an alternative route, enabling high-dimensional, parallel information processing directly in the physical layer, particularly suited for imaging tasks. In this context, quantum photonic platforms provide both a natural mechanism for computing inner products and a promising path to energy-efficient inference in photon-limited regimes.  
Here, we experimentally demonstrate a camera-free quantum-optical images classifier that performs inference directly at the measurement layer using Hong–Ou–Mandel (HOM) interference of spatially programmable single photons.
Two-photon coincidences directly report the overlap between an input image mode and a learned template, replacing pixel-resolved acquisition with a single global measurement.
We realize both a single-perceptron quantum optical neuron and a two-neuron shallow network, achieving high accuracy on benchmark datasets with strong robustness to experimental noise and minimal hardware complexity.
With a fixed measurement budget, performance remains insensitive to input resolution, demonstrating intrinsic robustness to the number of pixels, which would be impossible in a classical framework.
This approach paves the way toward neuromorphic quantum photonic processors capable of extracting task-relevant information directly from HOM interference, with promising applications in remote object recognition, low-signal sensing, and photon-starved biological microscopy.



\end{abstract}

\maketitle

\section{Introduction}

The past decade has witnessed unprecedented advancements in Machine Learning (ML) \cite{lecun2015deep, dale2025ai4x}, yielding outstanding results across different domains ranging from natural language processing to computer vision \cite{lecun2002gradient, krizhevsky2012imagenet, he2016deep}. 
Despite the exceptional success of this field, the rapid growth in model size and data dimensionality is exposing practical bottleneck of current hardware. This has motivated interest in alternative computational paradigms that can either overcome these limitations or potentially extend the current learning capabilities beyond standard silicon electronic processors.

Consequently, extensive research has been dedicated to implementing the resource-intensive computation required by neural networks.
On the classical side, this saw a renewed interest in neuromorphic computing \cite{shastri2021photonics, kudithipudi2025neuromorphic, brunner2025roadmap, li2025photonics} that exploits analog processors to perform multiply-accumulate operations with high parallelism to reduce the energy costs that dominate current ML workloads.
{Within this landscape, reservoir computing \cite{mujal2021opportunities, innocenti2023potential} has emerged as a particularly attractive framework, as it leverages the complex dynamics of physical systems to process information while requiring training only at the final linear readout stage. Recent experimental realizations in photonic platforms further underscore the promise of this approach for scalable and resource-efficient information processing \cite{cimini2025large, suprano2024experimental, zia2025quantum, di2025time}. This perspective naturally connects to the rapid progress in
the field of Quantum Machine Learning (QML) \cite{biamonte2017quantum, cerezo2022challenges, chang2025primer, zaman2023survey,  dunjko2018machine}, whose aim is to exploit quantum phenomena, such as superposition and entanglement, to capture complex data correlations otherwise inaccessible to classical models. In this context, a critical asset provided by quantum systems is the exponentially large dimensionality of Hilbert spaces with the number of information carriers, which can be harnessed for the efficient encoding and processing of high-dimensional data \cite{lloyd2013quantum, cai2015entanglement, cimini2025large, gong2025enhanced}. Within this framework, efficiently accessing inner products, within such spaces, becomes a central task that represents the core learning primitive, for example for kernel methods \cite{rebentrost2014quantum, havlivcek2019supervised, schuld2019quantum, yin2025experimental, hoch2025quantum}.
However, establishing when and why such approaches deliver a practical advantage is far from obvious, and it is challenging to generalize.

Photonic systems are a particularly favorable platform \cite{vernuccio2022artificial, sanchez2024advances}. Optics supports massive parallel processing and inherently high bandwidth, making this solution attractive, especially in terms of energy-efficient processing \cite{caulfield2010future, shastri2021photonics, mcmahon2023physics, fu2024optical, pai2023experimentally, zhou2022photonic, soret2026quantum}. This match is especially strong for optical image classification, where the raw input is an optical wavefront distributed over many spatial modes \cite{lin2018all}, and photonic platforms enable native processing of high-dimensional information encoded in optical modes \cite{ortolano2023quantum}. In conventional pipelines, this is first measured, converted into a pixelated electronic signal, and only then processed digitally. As a consequence, performance becomes fundamentally constrained on one hand by the characteristics of the measurement device (e.g. the number of pixels, readout noise, frame rate) and on the other by the available photon budget. In photon-starved or high-speed scenarios, these constraints are not merely technical, but they can dominate the achievable accuracy, latency, and robustness \cite{kwon2022bleaching}.
This point is particularly sharp in binary hypothesis testing problems that prevail in many practical imaging tasks, for example, deciding whether a specific marker, defect, or target signature is present or absent \cite{tufail2020binary, zerouaoui2022deep}. In such settings, the goal is not to reconstruct a detailed image, but to reliably discriminate between two alternatives often under tight photon budgets, limited exposure times, or power constraints. In such a framework, a camera-based approach is often suboptimal.
A more promising route is therefore to classify the image directly in the optical domain by extracting decision-relevant features from the optical system.
Many classifiers indeed reduce inference to a similarity test, which is based on the computation of an inner product or distance between an input vector and a matrix of learned weights in the high-dimensional feature space. Optics provides a natural mechanism for this operation by exploiting interference that effectively realizes the required operation in an analog way.
However, classical interferometric readouts often require stringent active phase stabilization routines that rely on more complex setups and electronics.

In this paper, to overcome this limitation, we exploit one of the building blocks of quantum optics effects: the Hong-Ou-Mandel (HOM) interference \cite{hong1987measurement,bouchard2020two, hiekkamaki2021high, jaouni2025tutorial, zhan2025experimental} that has been revealed as an especially powerful solution to address image classification tasks \cite{bowie2023quantum, roncallo2025qon, roncallo2025shallow}. Since we only need to recover a single bit (whether or not the object belongs to a certain class), we do not need to reconstruct the whole image, which would require a number of photons that scales as the image resolution.
Ideally, a single bit can be extracted from few photons, and it is encoded into the HOM quantum interference between input and probe photons, carrying on their respective spatial profiles the information relative to the image to be classified and the model parameters tuned during the training stage. Conceptually, this implements the core operation of a classical perceptron in a resource-efficient way by computing distances within the feature space \cite{bowie2023quantum}, with the practical advantage of a minimal experimental setup. In more detail, in the HOM-based quantum optical neuron (QON), the input is encoded in the single-photon state while the trainable parameters are encoded in a second single-photon state. After HOM interference, the two-photon coincidence probability depends on the distinguishability of these states and directly yields a similarity score proportional to the squared overlap. Crucially, in this scenario, the readout is performed with two bucket detectors without spatial sensitivity, one per output port, so the classifier output is obtained from the coincidence rate rather than from a pixel-resolved measurement. 
This measurement choice directly targets the two bottlenecks that limit camera-based inference. First, it is robust against scaling with the number of pixels, and secondly, it is naturally suited to photon-starved regimes. When photons are scarce, distributing them across many pixels makes individual intensity estimates noisy and forces longer integration times. By contrast, HOM readout concentrates the task into estimating a single probability, so the photon budget is spent on the decision variable itself rather than on reconstructing an intermediate image.
Here, we report the first experimental realization of this HOM-based QON and its extension as quantum optical shallow network (QOSN) with two neurons, demonstrating high performance in binary image classification benchmark tasks and strong robustness to the input image dimensionality.

 Our results establish HOM interference as a practical primitive for a regime where imaging is hardest, few photons, short exposures, and weak returns, opening a route toward neuromorphic quantum photonic processors making reliable decisions where image reconstruction is unnecessary.
This approach paves the way to efficient image processing across numerous applications, ranging from remote object recognition to photon-starved biological microscopy, in which the information to be manipulated is natively encoded in an optical field.

\begin{figure}[ht]
    \centering
    \includegraphics[width=\linewidth]{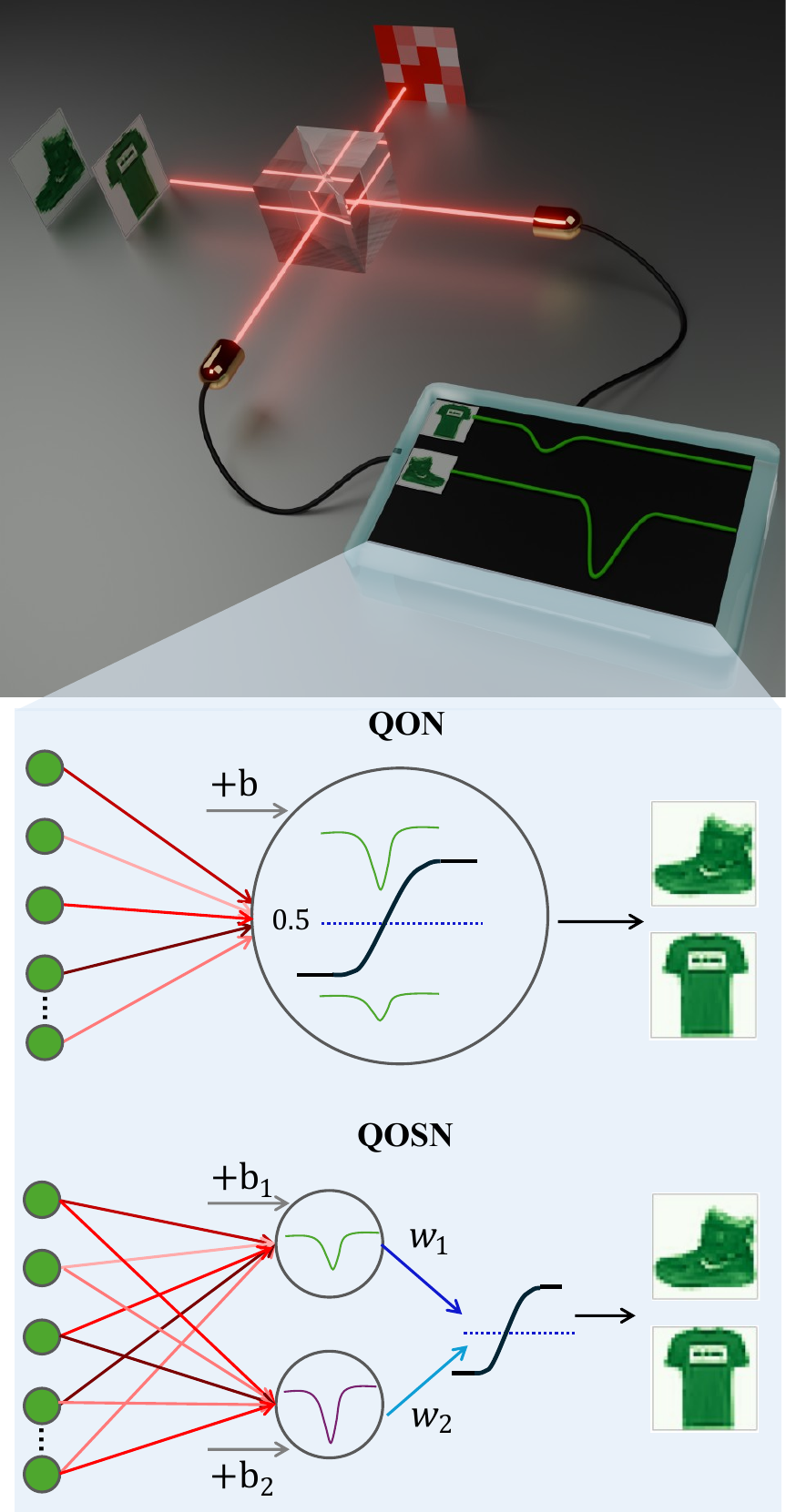}
    \caption{\textbf{QON and QOSN scheme.} The core of the QON operation is Hong-Ou-Mandel (HOM) interference between two spatially modulated photons. The input photon (green profile) carries the spatial information of the image to be classified, while the probe photon encodes the model weights. The neuron's weighted sum is physically implemented via the HOM interference. Bucket detectors provide the HOM visibility, which, combined with a trainable bias, serves as the input for a sigmoid activation function predicting the input image's class.}
    \label{fig:concept}
\end{figure}

\section{Results}
\subsection{Theoretical framework} 
\label{sec:theo}

We review how to use the Hong-Ou-Mandel effect to implement artificial neurons and shallow neural networks on an optical device. Consider two photons separately fed into a balanced beam splitter. We encode the input data $X$ and the model weights $\lambda$ in the spectral amplitudes of the photon spatial modes (in the monochromatic approximation), through a pure state and a density operator $\rho$, respectively. The detection is performed by two bucket detectors with no spatial resolution, placed after the beam splitter, which click whenever a photon exits their mode. Due to the Hong-Ou-Mandel effect, the coincidence rate $p_c$ at the output - the probability that both detectors click - depends on the spatial-spectrum distinguishability $z(X)$ between the two input modes. Namely, $p_c = [C - z(X)]/2$, where $C$ is a constant depending on the normalization of the spectrum but also on the presence of photon losses in the apparatus. Denoting $U$ the spatial modes of $\rho$, one can show that 
\begin{equation}
	z(X) = \int \text{d}k_1 \text{d}k_2 X(k_1)X^*(k_2)U_\lambda(k_2,k_1) \ ,
\end{equation}
with $X(k)$ the Fourier spectrum of the photon encoding the input data, and $k$ the spatial momentum on the image plane (where the beam splitter is placed). A complete derivation of this equation is discussed using second quantization and Wick's theorem in \cite{roncallo2025qon,roncallo2025shallow}.

For a mixture of $M$ pure states with spectrum $W_{\lambda_i}$ and probability $w_i$, we have $U_\lambda(k_2,k_1) = \sum_i w_i W_{\lambda_i}(k_2)W^*_{\lambda_i}(k_1)$, where $i=0,\ldots,M-1$. By repeating the measurement, one obtains $z(X)$ from the coincidence rate. We then add a bias $b$ and a sigmoid activation function $\sigma(t) = 1/[1+\exp(-t)]$ in post-processing. This gives
\begin{gather}
        \label{eq:act_func}
	f_\theta(X) = \sigma(z(X)+b) \ , \\
	z(X) = \sum_{i=0}^{M-1}w_i|\langle X, W_{\lambda_i} \rangle|^2 \ .
	\label{eq:Network}
\end{gather}
Here $\langle \cdot, \cdot \rangle$ denotes the $L^2$ inner product of the Hilbert space, which due to the Plancherel's theorem can be either computed in the position or in the momentum representations, and $\theta \in \{w,W_\lambda,b\}$ the set of trainable parameters. For discrete spatial modes, this equation reproduces the output of a classical shallow neural network, i.e.~a single layer of $M$ hidden neurons with square absolute value activations followed by an output neuron with a bias and a sigmoid activation. Our model exhibits additional constraints imposed by the $L^2$ and $L^1$ normalization, i.e.~that $\| W_{\lambda_i} \|^2 = 1 \ \forall i$, and $\sum_i w_i = 1$. 

For binary classification tasks, consider the training set $D = \{X_s\}$ with known labels $\{y_s \in \{0,1\}\}$, e.g.~a collection of images of 0s and 1s. We train the parameters by comparing the predicted classes with the target labels, with the binary-cross entropy $H$ as loss function. This means solving the optimization problem $\min_{\theta} \sum_{s \in D} H(y_s, f_{\theta}(X_s))$. We optimize $H$ via gradient descent, where the parameters are updated by computing the gradients of the loss function as
\begin{equation}
    \theta \to \theta - \frac{\eta}{|D|} \sum_{s\in D} \nabla_\theta H\left(y_{s},f_{\theta}(X_s)\right) \ ,
    \label{eq:GradientDescent}
\end{equation}
with $\eta$ called learning rate. A detailed calculation of the gradients $\nabla_\theta H$ both for the single neuron and the shallow network are reported in \cite{roncallo2025qon,roncallo2025shallow}. This update rule is applied at each epoch, i.e.~a full pass through the dataset. After training, the model can perform inference and generalize to new data, showing an advantage over its classical counterpart, as illustrated below.

We discuss the resource cost of our protocol and the quantum advantage. A neural network with $N$ input neurons and $M$ hidden neurons has a computational cost that scales at least linearly in the number of mathematical operations, i.e.~$\mathcal{O}(MN)$ resources in inference ($M$ scalar products of $N$-dimensional arrays). When this is applied to images, an additional optical cost arises due to image acquisition. In the best case scenario, retrieving a digital black-and-white image of $N$ pixels requires $\mathcal{O}(N)$ photons. Using the HOM effect, we implement the same network, but optically (without classical hardware) and using only bucket detectors (without image retrieval). Such computation is not deterministic: estimating the coincidence rate requires multiple repetitions, i.e.~photons counts. For the $i$th pure state, this is equivalent to retrieving a binomial distribution from its empirical frequencies with uncertainty $\varsigma$ (with success-failure corresponding to the presence-absence of coincidence), i.e.~a constant overhead of $\mathcal{O}(\varsigma^{-2})$ photons. Due to Hoeffding’s inequality \cite{hoeffding1963probability}, this holds true also when sampling from a mixture of $M$ pure states. In this case, estimating the coincidence rate is equivalent to taking the sample average of a random variable with outcomes $|\langle X, W_{\lambda_i} \rangle|^2$ and probability $w_i$. Namely, estimating $p_c$ with uncertainty $\varepsilon$ and confidence level $1-\delta$ requires $\mathcal{O}(\varepsilon^{-2}\log(2/\delta))$ photons. In conclusion, both methods achieve inference with constant $\mathcal{O}(1)$ photons and mathematical operations, i.e.~independently of the input resolution. A rigorous derivation of the advantage is discussed in the supplementary material of \cite{roncallo2025qon,roncallo2025shallow}.

\subsection{Experimental implementation}
\label{sec:exp_setup}

\begin{figure*}[ht!]
    \centering
    \includegraphics[width=\textwidth]{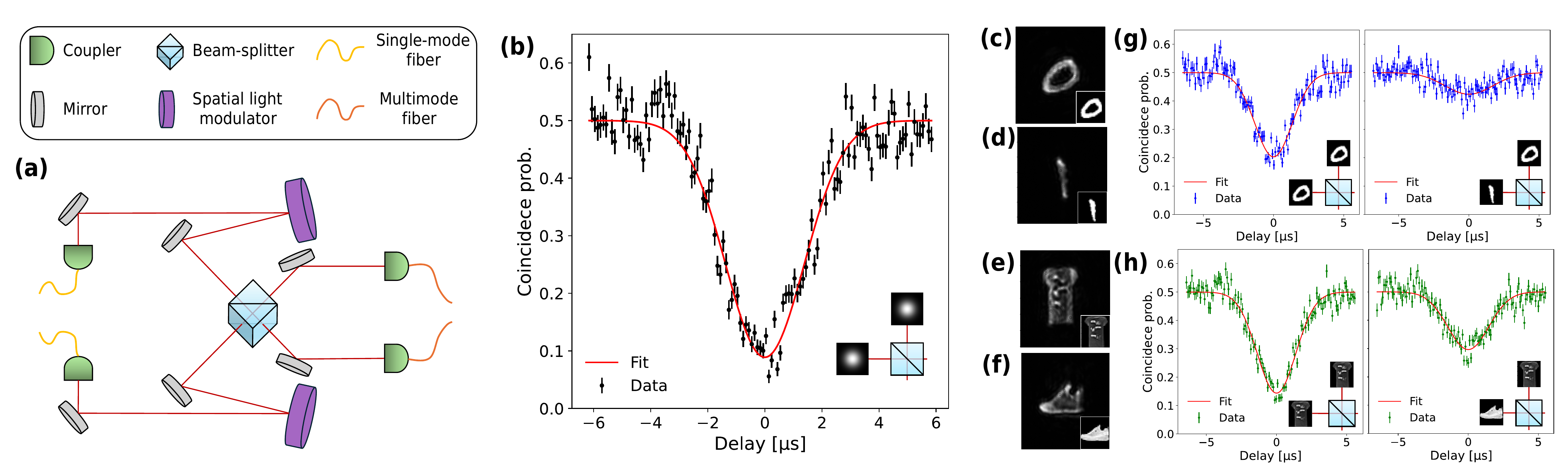}
    \caption{\textbf{Experimental setup characterization.} Panel \textbf{(a)} illustrates the schematic of the experimental setup used to implement the QON. Pairs of single photons, initialized in the $TEM_{00}$ spatial mode, are modulated by spatial light modulators (SLMs) to generate arbitrary transverse amplitude profiles at the beam-splitter plane. Following interference at the beam-splitter, the photons are coupled into multimode fibers to measure coincidence counts and, ultimately, determine the Hong-Ou-Mandel (HOM) visibility. In panels \textbf{(b)}-\textbf{(h)}, we report the characterization of the HOM interference for different spatial profiles, obtained by measuring coincidence counts as a function of the relative time delay between photons. In panel \textbf{(b)}, we consider the case of Gaussian profiles (illustrated in the inset), which yield an HOM visibility of ($84.5 \pm 2.4$)\%. Panels \textbf{(c)}-\textbf{(d)} and \textbf{(e)}-\textbf{(f)} show the results of spatial modulation applied to a coherent beam for example images from the MNIST and Fashion-MNIST datasets, respectively, with target images shown in the insets. Panel \textbf{(g)} displays the HOM dips resulting from the combinations of MNIST ``0-0'' and ``0-1'' images, while panel \textbf{(h)} considers combinations of ``t-shirt--t-shirt'' and ``t-shirt--sneaker'' from the Fashion-MNIST dataset.}
    \label{fig:setup}
\end{figure*}

The realization of the QON and QOSN 
relies on a single-photon source that generates photon pairs via spontaneous parametric down-conversion (SPDC), programmable spatial mode preparations, and interference on a bulk 50:50 beam-splitter (BS).
A schematic of the experimental setup is illustrated in Fig.~\ref{fig:setup}a.
The SPDC source provides pairs of single photons that are made indistinguishable in polarization, frequency spectrum, arrival time, and spatial mode before being injected into the QON setup through single-mode fibers (additional details on the single-photon source are provided in the Materials and Methods section).

The two photons are routed into the two arms of the setup and act respectively as the input photon, carrying the image to be classified, and the probe photon, whose profile is trained during the algorithm running.
The image we want to classify is encoded by imprinting a programmable mask on the amplitude profile of the input photon via a spatial light modulator (SLM). In the second arm, an independent SLM prepares the amplitude profile of the probe photon, implementing the trainable weights of the neuron. More details on the programmable masks are reported in the Materials and Methods section.


Specifically, we prepare the input photon state to reproduce the image to be classified and the probe photon to encode the weights, both within their amplitude profiles at the beam-splitter plane, where the two photons interfere, while maintaining a flat phase. 

As described in Sec.\ref{sec:theo}, this encoding choice ensures that the theoretical model exactly matches the action of an artificial neuron. The spatial overlap between the SLM-modulated profiles of the two photons determines the strength of the HOM interference, which is readily identified from the detection coincidence rate of the two final bucket detectors. Experimentally, these are implemented by coupling the incoming photons into multimode fibers (MMF) connected to Avalanche Photodiodes (APD). The estimated HOM visibility $z_{\mathrm{exp}}$ is derived from the fractional drop in coincidence counts, $z_{\mathrm{exp}} = 1 - C_{\mathrm{min}} / C_{\mathrm{max}}$, resulting from the interference between the weight-shaped photon and the sample-shaped photon. Following Eq.\eqref{eq:act_func}, this quantity is then post-processed using a biased sigmoid function $\sigma(z_{\mathrm{exp}} + b)$, whose output determines the model's prediction.

This method provides a suitable framework to engineer the single-photon spatial profiles, and to readily measure the corresponding HOM visibility. In contrast, the mixed states required to implement the QOSN model are obtained by post-processing the visibilities provided by individual probe profiles, effectively yielding the output described in Eq.\eqref{eq:Network}.
We highlight that the MMF-based detection scheme employed in this experiment eliminates spatial resolution, and it offers significant benefits in terms of both protocol complexity and statistical noise resilience. Indeed, it enables straightforward post-processing of the measured data, as the HOM visibility is directly fed into the activation function. Simultaneously, by integrating over many spatial modes, this detection scheme dramatically lowers the experimental requirements for achieving sufficient statistical significance in the measured coincidence counts (see Supplementary Information).

To benchmark the performance of our experiment and validate the key resource exploited in this work, we first characterized the HOM interference for a range of spatial profiles of the signal and idler photons. By measuring the coincidence counts as a function of the relative time delay between the photons, we observe the characteristic HOM dip and extract the corresponding interference visibility. In particular, in Fig.\ref{fig:setup}b, we consider the simplest case in which both the profiles are kept in the fundamental gaussian mode, achieving a HOM visibility $v = (84.5 \pm 2.4)\%$. This condition is obtained by leaving flat phase masks on the SLMs, and it sets the reference for the maximum indistinguishability achievable by our source. 

For our protocol, we also have to consider the effects of HOM interference between spatially engineered single-photon states. In this scenario, the relevant figure of merit is no longer a fixed source visibility, but the visibility associated with the mode overlap between the prepared profiles.
To this end, we program the SLM phase masks to generate structured spatial amplitudes corresponding to representative samples from the ``0''/``1'' images of the MNIST handwritten-digits dataset and ``t-shirt/sneaker'' images of the fashion-MNIST dataset. 
Before operating at the single-photon level, we verify the correctness of each programmed mask by injecting a bright coherent beam along the same optical paths and measuring the resulting intensity distribution with a beam profiler, reported in Fig.\ref{fig:setup}c-f, ensuring faithful reproduction of the intended spatial patterns at the interferometer input.
In our detection configuration, the HOM visibility provides a direct experimental proxy for the spatial overlap between the quantum state of the input and probe photons \cite{jaouni2025tutorial}. Fig.\ref{fig:setup}g-h, show, in the cases of MNIST and fashion-MNIST sample images respectively, how higher (lower) visibilities are obtained when the spatial profiles of the two photons are identical (different). This visibility gap is found to be larger for the MNIST samples than for the Fashion-MNIST ones, serving as an indirect manifestation of the relative classification difficulty associated with each dataset.
This immediate connection between visibility and spatial overlap is at the core of the QON and QOSN models, which we explore further in Sec.~\ref{sec:results} through the analysis of the experimental data.

\subsection{Images classification performance}
\label{sec:results}

\begin{figure*}[ht!]
    \centering
    \includegraphics[width=\linewidth]{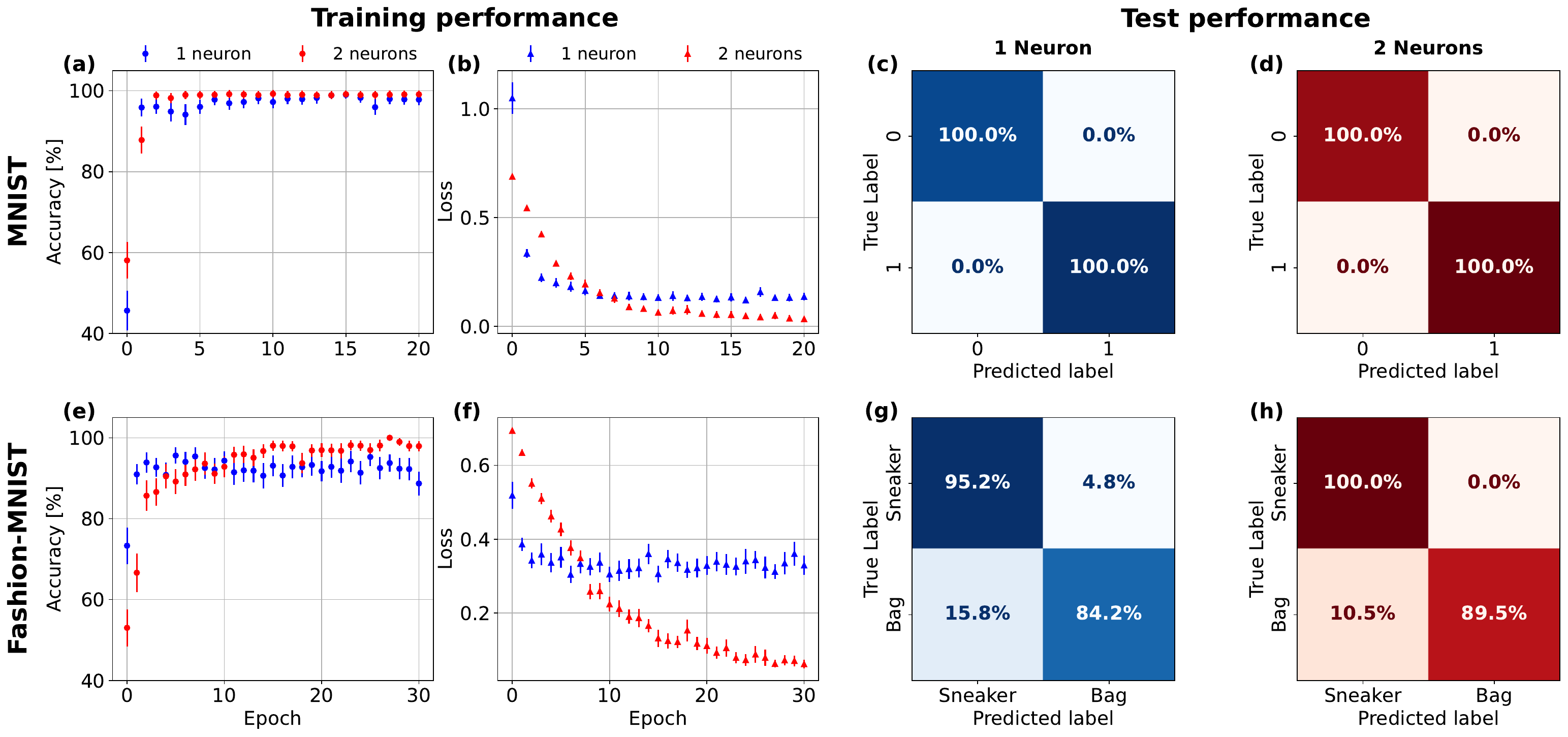}
    \caption{\textbf{QON and QOSN performance.} Panels \textbf{(a)}-\textbf{(b)} and \textbf{(e)}-\textbf{(f)} display the evolution of training accuracy and loss over the training epochs. Panels \textbf{(a)}-\textbf{(b)} correspond to the classification of MNIST ``0'' vs.\ ``1'' images (20 epochs), while panels \textbf{(e)}-\textbf{(f)} correspond to Fashion-MNIST ``sneaker'' vs.\ ``bag'' images (30 epochs). Blue markers represent the single QON performance, while red markers indicate the results obtained with a two-neuron QOSN. The errorbars reported in panels \textbf{(a)}-\textbf{(b)} and \textbf{(e)}-\textbf{(f)} correspond to the standard deviation achieved on 100 independent random choices of 50-sample subsets of the training set onto which the accuracies and loss have been evaluated. Panels \textbf{(c)}-\textbf{(d)} and \textbf{(g)}-\textbf{(h)} report the confusion matrices evaluated on the test datasets for MNIST and Fashion-MNIST, respectively. In particular, panels \textbf{(c)} and \textbf{(g)} show the results for the single QON, while panels \textbf{(d)} and \textbf{(h)} correspond to the two-neuron QOSN.}
    \label{fig:train_test}
\end{figure*}

In what follows, we report the experimentally achieved classification performance of the model described in Sec.~\ref{sec:exp_setup}, considering the cases of the single QON and the 2-neuron QOSN. For this purpose, we employed two benchmark datasets of increasing complexity, the MNIST \cite{lecun2002gradient} and Fashion-MNIST \cite{xiao2017fashion} datasets, respectively. The dataset examples consist of $32\times32$ grayscale images ($28 \times 28$ images plus a 2-pixel padding on each side) representing handwritten digits for the MNIST dataset and clothing items for Fashion-MNIST. Since both the QON and QOSN models are designed to perform binary classification, we constructed training (test) sets by randomly sampling 100 (40) images, restricting the datasets to the ``0'' and ``1'' classes for MNIST, and the ``sneaker'' and ``bag'' classes for Fashion-MNIST. The neuron weights encoded in the amplitude profile of the probe photon, and, in the case of the QOSN, the two output weights, are initialized by sampling from a uniform distribution between 0 and 1, while the bias is initialized to 0 (more details are reported in the Materials and Methods section). Training then proceeds for 20 and 30 epochs for the MNIST and Fashion-MNIST classification tasks, respectively, with the learning rate decreased at the midpoint of the training process. At each epoch, the model parameters are updated via gradient descent based on the experimentally estimated HOM visibilities on the overall training dataset and, consequently, the SLM phase mask of the probe photon is modified to physically reproduce this newly updated set of weights. 

In Fig.\ref{fig:train_test}, we compare the training classification performance of the single QON and the two-neuron QOSN on both datasets as a function of the training epochs. The evolution of accuracy and loss indicates that both models reach asymptotic performance within the allotted training epochs. Specifically, at the final epoch, the classification accuracies reach $97.8 \pm 1.4$ ($88.7\pm 3.0$) for the QON and $99.1\pm 1.0$ ($97.9\pm 1.3$) for the QOSN on the MNIST (Fashion-MNIST) dataset. 
The performance gap is modest for MNIST but becomes pronounced for Fashion-MNIST, consistent with the increased complexity of the Fashion-MNIST discrimination. The classification performance achieved on independent test sets validates this trend, as indicated by the accuracies reported in the form of confusion matrices in Fig.\ref{fig:train_test}(c)-(g) and Fig.\ref{fig:train_test}(d)-(h) for the QON and QOSN results, respectively. Using the weights learned during the training, both models achieve 100\% classification accuracy on the MNIST test images, while on the Fashion-MNIST test dataset, the QON reaches 90\% and the QOSN 95\%. Together, these results suggest that, as expected, the two-neuron shallow network outperforms the single neuron, with an enhancement that becomes more evident as the classification task becomes more complicated.

\begin{figure}[ht]
    \centering
    \includegraphics[width=\linewidth]{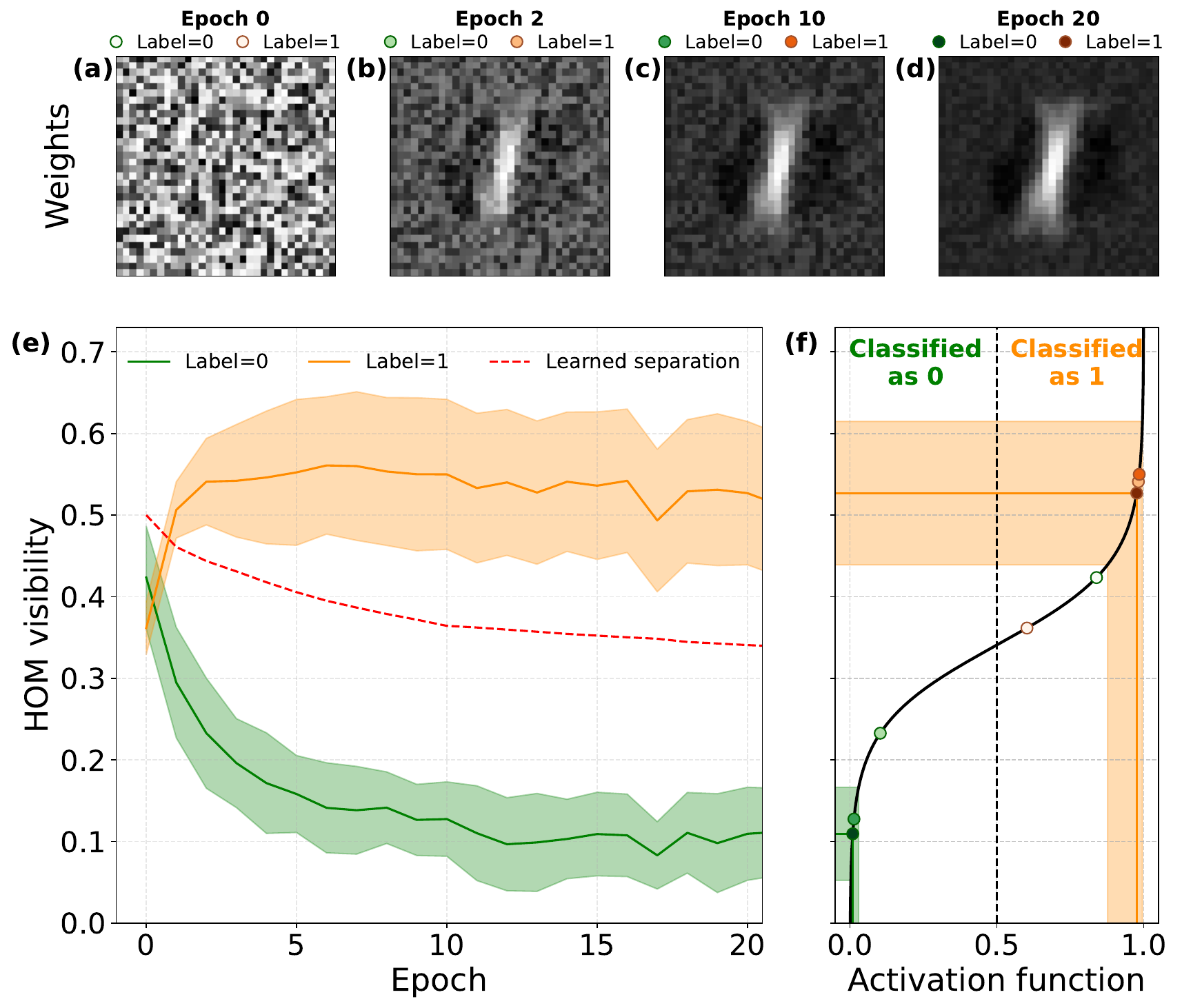}
    \caption{\textbf{HOM visibilities during the model training} Panels \textbf{(a)}-\textbf{(d)} illustrate representative examples of QON weights during the training, reporting, in particular, their profiles at epochs 0, 2, 10, and 20. In panel \textbf{(e)}, we report the evolution of the measured HOM visibilities averaged on the train dataset, separated for instances of MNIST images of zeros (green solid line and shaded area) and ones (orange solid line and shaded area). The red dashed line represents the learned threshold for which the model performs the binary classification, obtained as 0.5-$b_i$, where $b_i$ represents the current epoch bias. Panel \textbf{(f)} illustrates the correspondence between the final average and standard deviation of the visibilities for the two classes and the final prediction obtained as the output of the sigmoid activation function (black solid line). Dots in different shades of green and orange illustrate the average biased visibilities of ``0'' and ``1'' MNIST samples, respectively, at epochs 0, 2, 10, and 20.}
    \label{fig:vis_train}
\end{figure}

To gain deeper insight into what training is doing, we examine how, for different epochs, the different trained weights masks impact the physical quantity, which is actually measured, namely, the HOM visibility. Fig.\ref{fig:vis_train} illustrates this for the binary classification of MNIST images. In Fig.\ref{fig:vis_train}a-d, it is illustrated how the training process converges toward a weight profile with higher (lower) amplitudes in regions typically occupied by high-intensity pixels of ``1'' (``0'') images. As depicted in Fig.\ref{fig:vis_train}e, this dynamic effectively maximizes the visibility gap resulting from the HOM interference between single photons whose spatial profiles encode the weights and those encoding the ``0''/``1'' MNIST images. At the same time, the bias is trained to find the optimal separation boundary between the visibilities corresponding to different classes. Fig.\ref{fig:vis_train}f shows indeed that this separation effectively produces, via the post-processing provided by the sigmoid activation function, the final neuron outputs that correctly predict the images class of belonging. This physical interpretation of the training process also highlights the remarkable noise resilience of the QON and QOSN models, as good classification performances can be achieved even when noise limits the maximum achievable HOM visibility and, consequently, the visibility gap obtained during training (further details are provided in the Supplementary Information).

Finally, we studied the protocol's robustness against the input resolution. We repeat the classification experiment for dataset where we progressively change the number of pixels used to represent each image example, while keeping all the other resources and settings fixed. In particular, we do not change the experimental budget used to estimate the HOM visibility (same number of coincidence counts per data point), nor the training procedure and hyperparameters. In these conditions, we observe that the classification performance remains essentially unchanged as the resolution is reduced, within statistical fluctuations. In contrast to camera-based approaches, where both acquisition and noise typically scale with the pixel counts, here, increasing the input dimensionality does not alter the measurement and processing structure that both rely only on the measure of the HOM visibility. 

Clearly, in order to do image classification in a classical setting, one has to first recover the image. In the HOM setting we detail here, instead, a single bit is extracted from the protocol (whether or not the object belongs to a class), and this can be attained with a limited number of repetitions of the protocol, that does not depend on the resolution of the image that has to be classified. 

\begin{figure*}[ht]
    \centering
    \includegraphics[width=\linewidth]{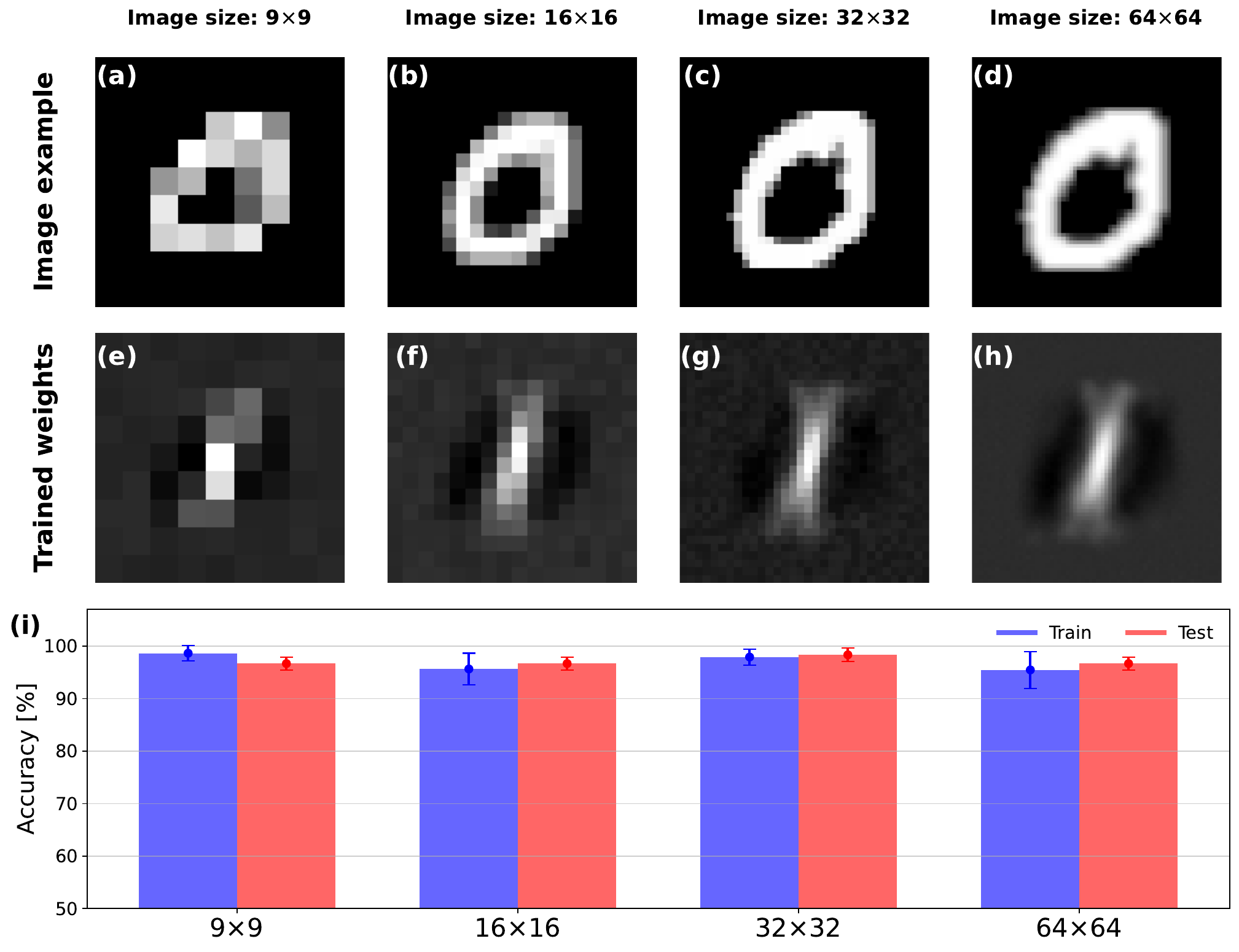}
    \caption{\textbf{QON performances with variable resolution.} Panels \textbf{(a)}-\textbf{(d)} illustrate a representative example of an MNIST image at different resolutions, i.e., $9\times9$, $16\times16$, $32\times32$, and $64\times64$ pixels, while the central row of panels \textbf{(e)}-\textbf{(h)} shows the corresponding trained weights profiles. In panel\textbf{(i)}, we report the training and test accuracies achieved using image datasets of varying resolutions. The error bars represent half the maximum variation observed over the final 10 epochs, capturing the fluctuations after the accuracies have converged.}
    \label{fig:ris_variabile}
\end{figure*}


\section{Discussion}
In this work, we present the experimental realization of both the QON \cite{roncallo2025qon} and QOSN \cite{roncallo2025shallow} models, achieved by exploiting the HOM interference of spatially modulated single-photon pairs. The full parallelization of scalar product operations, analogous to the action of a single perceptron or a multi-neuron shallow network, enables us to process and classify images while entirely bypassing the need to digitize them. Consequently, this approach incurs only a constant resource cost, effectively eliminating the standard restrictions on the resolution of the images to be classified.
We benchmarked the learning capability of these models on binary image-classification tasks across standard datasets, observing rapid convergence and high accuracy, with a clear improvement when moving from a single QON to a two-neuron QOSN on more challenging datasets.
In addition to the optically enabled parallelization, another pivotal feature of the QON setup is its minimal complexity, which offers a twofold advantage. First, it significantly relaxes the experimental requirements on the photon generation rate, as the signal does not need to be partitioned across multiple pixels, while, at the same time, it renders the detection and post-processing stages unaffected by scaling in image resolution.

While this work serves as a proof-of-principle to benchmark the capabilities of the QON and QOSN models, the presented approach targets a broad class of inference problems in which information is natively carried by an optical field and the objective is decision-making rather than image reconstruction.
A particularly natural domain of applicability is binary hypothesis testing in imaging, where the goal is to decide whether a marker, anomaly, or target signature is present rather than to form a high-fidelity image. Examples include biological marker detection and remote recognition tasks where eye-safe illumination, specimen safety, weak returns, or short exposure times impose strict photon budgets. In these regimes, pixel-resolved imaging can be fundamentally inefficient because it spends photons estimating many degrees of freedom that are irrelevant to the final decision.
The QON  scheme is therefore well suited to remote sensing and imaging protocols where HOM interference has already been proven to represent a great resource \cite{ndagano2022quantum, bornman2019ghost}, and our results provide a concrete experimental route to integrate HOM-based measurements into learning-enabled decision systems.



\section{Materials and methods}
\subsection{Single-photon source}
\begin{figure}[ht]
    \centering
    \includegraphics[width=\linewidth]{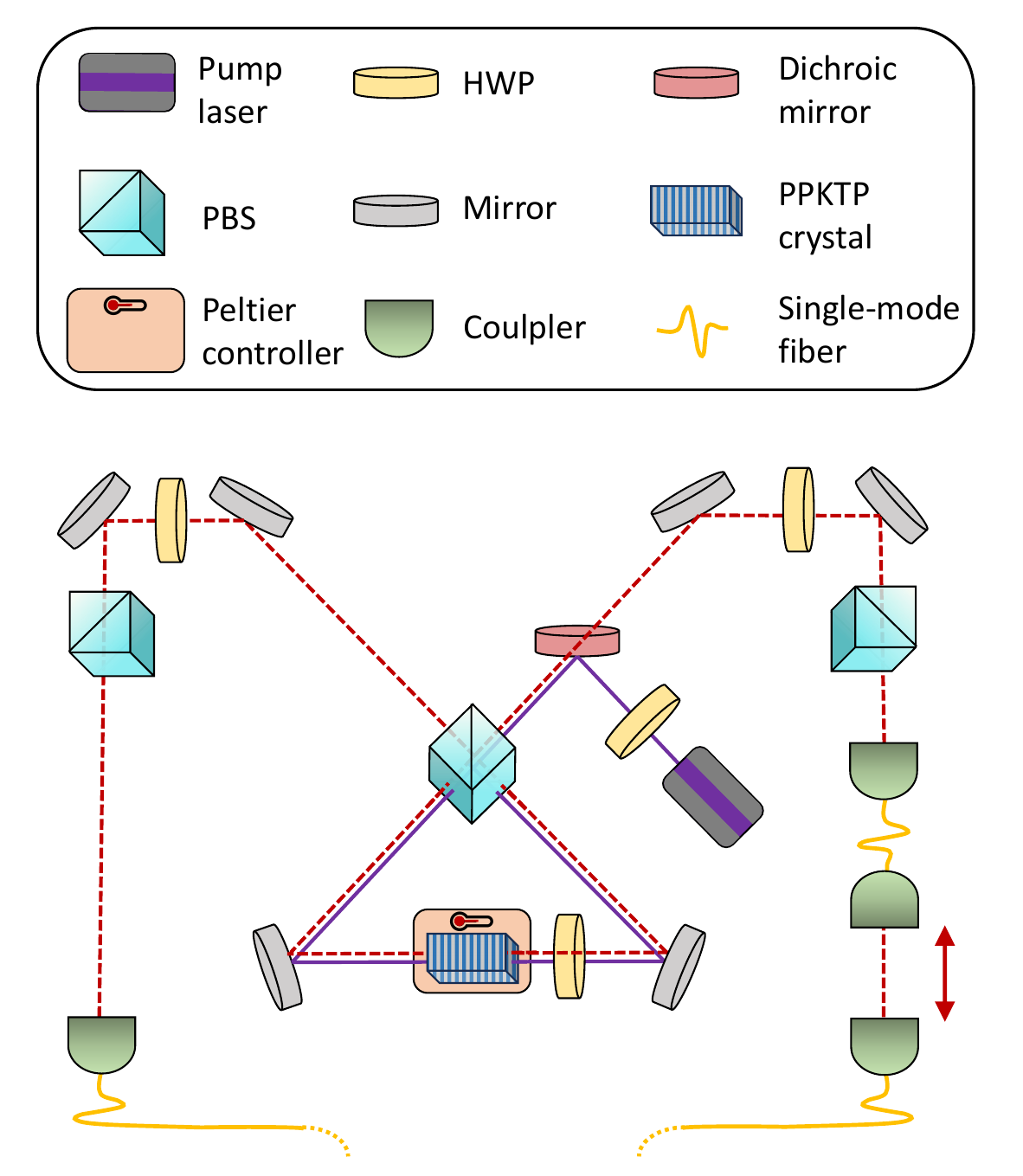}
    \caption{\textbf{Single-photon pair source.} A 404 nm horizontally polarized pump laser is injected into a Sagnac interferometer via a Polarizing Beam Splitter (PBS) to induce Spontaneous Parametric Down-Conversion (SPDC) in a type-II Periodically Poled Potassium Titanyl Phosphate (PPKTP) crystal, generating 808 nm single-photon pairs. The crystal temperature, and thus the emission spectrum, is stabilized using a Peltier controller. Half-Wave Plates (HWPs) and additional PBSs are used to ensure strictly horizontal polarization of the output photons. Finally, a motorized stage controls the relative time delay between the photons before they are routed to the main QON setup via single-mode fibers.}
    \label{fig:sagnac}
\end{figure}
Single-photon pairs are generated within a Sagnac configuration in which a periodically poled potassium titanyl phosphate (ppKTP) nonlinear crystal is pumped at 404 nm, thereby producing single photons at 808 nm via spontaneous parametric down-conversion (SPDC). By setting the pump to be horizontally polarized, the photon pairs are generated in the separable polarization state $\ket{H}_i\ket{V}_s$, where the two photons are labeled as \textit{idler} and \textit{signal}, denoted by the subscripts $i$ and $s$, respectively. Before sending the two photons to the QON setup, they must be made indistinguishable across all their degrees of freedom. For this purpose, we change the signal polarization to match that of the idler using a half-wave plate (HWP), while the spectral overlap and relative time delay are optimized by adjusting the crystal temperature and the length of a delay line on the signal path. The resulting photon pairs are then sent to the QON setup via single-mode fibers.

\subsection{Spatial Light Modulator masks}
\label{sec:slm_masks}
To reshape the spatial profile of input and probe photons into target patterns (MNIST and Fashion-MNIST images for the former, weights profile for the latter) via amplitude encoding, we employ a phase-only Spatial Light Modulator (SLM). In particular, we aim to modify an initial gaussian beam via a phase mask that, upon propagation up to the BS, reproduce the target image in the amplitude profile, while the phase remains flat. To achieve this, we generate the corresponding holograms using a modified Gerchberg-Saxton iterative phase retrieval algorithm \cite{wu2015simultaneous}. 
\begin{figure}[ht]
    \centering
    \includegraphics[width=\linewidth]{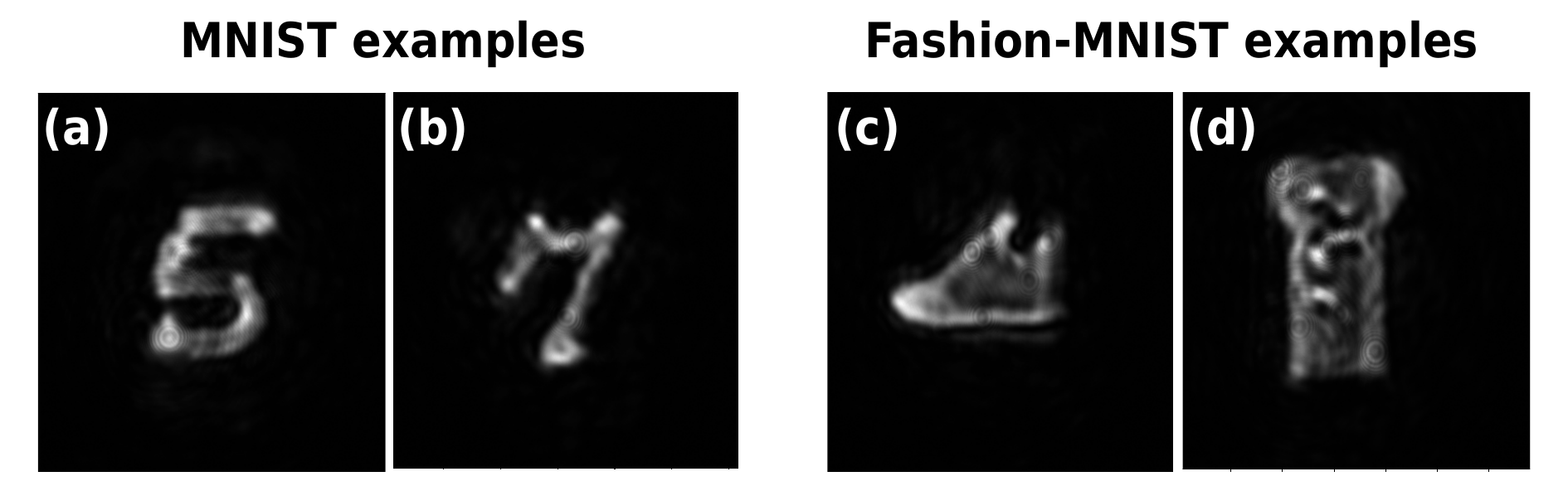}
    \caption{\textbf{Examples of experimental profiles.} In panels \textbf{(a)}-\textbf{(b)} and \textbf{(c)}-\textbf{(d)}, we respectively illustrate examples of MNIST and fashion-MNIST reconstructed onto the amplitude profile with the method described in \cite{wu2015simultaneous}. In this case, the SLM phase masks have been applied to a coherent beam, allowing to capture the final profiles via a CCD camera.}
    \label{fig:examples}
\end{figure}
This method leverages repeated free-space propagation and backpropagation to simultaneously shape both the amplitude and phase of the optical field. Computationally, the target output plane $(x, y)$ is mapped onto two completely overlapped, mutually complementary virtual planes, denoted as $\alpha$ and $\beta$. At any given coordinate, a point constrained on one plane must be unconstrained on the other.  We define the input field amplitude $a_t$, initial phase $p_0$, target amplitude $A_t$, target phase $P_t$, and the total number of iterations $N$. During each iteration, the input field $u_{\mathrm{in}} = a_t \cdot e^{i p_0}$ undergoes forward propagation $\mathcal{T}$ by a distance $L_z = z_{\mathrm{BS}} – z_{\mathrm{SLM}}$, yielding the field $U_c = \mathcal{T} [a_t \cdot e^{i p_0}] = A_c e^{i P_c}$. In particular, we define $\mathcal{T}$ following the angular spectrum method \cite{goodman1969introduction}. This field is then distributed across planes $\alpha$ and $\beta$. Using the unit matrix $I$ and a binary constraint matrix $S$ (with entries equal to 1 to indicate a constrained pixel and 0 an unconstrained one), the modified amplitudes and phases for each plane are computed as:
\begin{equation}
\left\{
\begin{aligned}
    A_\alpha &= A_t \cdot S + A_c \cdot (I - S)\\
    P_\alpha &= P_t \cdot S + P_c \cdot (I - S)\\
    A_\beta &= A_t \cdot (I - S) + A_c \cdot S\\
    P_\beta &= P_t \cdot (I - S) + P_c \cdot S
\end{aligned}
\right.
\end{equation}
Therefore, backpropagating these fields, we obtain $u_{\alpha,\beta} = \mathcal{T}^{-1}(A_{\alpha,\beta} \ e^{i P_{\alpha,\beta}}) \equiv a_{\alpha,\beta} \ e^{i p_{\alpha,\beta}}$. The phases of these backpropagated fields are combined to determine the updated input phase $p_0 \to p_1=p_\alpha + p_\beta$ for the subsequent iteration. After repeating this procedure for $N$ iterations, the phase mask is given by the final phase $p_N$.
In the experimental implementation of the QON and QOSN, we generated the phase masks by iterating 25 times the algorithm described above. Examples of MNIST and fashion-MNIST images reconstructed on the amplitude profile are illustrated in Fig. \ref{fig:examples}.

\subsection{Model hyperparameters and experimental details}
This section outlines the QON and QOSN hyperparameters used in our experimental implementation. For both models and datasets (MNIST and Fashion-MNIST), we utilize training and test sets of 100 and 40 randomly sampled images, respectively. While 20 epochs are sufficient to train the models for MNIST classification, we increase this to 30 epochs for Fashion-MNIST to ensure better convergence. Several learning rates are defined within the QON and QOSN model; specifically, $\eta_\lambda$, $\eta_b$, and $\eta_w$ govern the updates for the weights, biases, and (for the QOSN) the output weights, respectively. Rather than keeping these rates constant, we decrease them halfway through training to allow for fine-tuning. In the QON model, we reduce $\eta_b$ by a factor of 3 while keeping $\eta_\lambda$ fixed. Conversely, in the QOSN model, we halve all initial learning rates at the 10th epoch for MNIST and the 15th epoch for Fashion-MNIST. We also optimize the shape of the sigmoid activation function using the hyperparameters $\beta$ and $\gamma$, defined as $\sigma_{\beta \gamma}(t) = 1/[1 + \mathrm{exp}(- \beta t + \gamma)]$. For the QON model, these are set to $\beta=11$ and $\gamma=5.5$, respectively, while for the QOSN model, we fix them to $\beta=1$ and $\gamma=0$. Table \ref{tab:hyperpar} summarizes all model hyperparameters for both MNIST and Fashion-MNIST image classification across the QON and QOSN models. Additionally, the last row of the table reports the average signal measured in the experimental implementations, defined as the average number of detected coincidence counts in the absence of HOM interference.

\begin{table}[ht]
    \centering
    \label{tab:hyperpar}
    \begin{tabular}{c|cccc}
        
        & \multicolumn{2}{c}{\textbf{MNIST}} & \multicolumn{2}{c}{\textbf{Fashion-MNIST}} \\
        \toprule
         & \textbf{1 neuron} & \textbf{2 neuron} & \textbf{1 neuron} & \textbf{2 neuron} \\
        \hline
        $N_{\mathrm{train}}$ & 100 & 100 & 100 & 100 \\
        $N_{\mathrm{test}}$ & 40 & 40 & 40 & 40 \\
        $N_{\mathrm{epoch}}$ & 20 & 20 & 30 & 30 \\
        $\eta_\lambda$ & 0.075 & 0.3 (0.15) & 0.075 & 2.1 (1.05) \\
        $\eta_b$ & 0.015 (0.005) & 0.6 (0.3) & 0.015 (0.005) & 4.2 (2.1) \\
        $\eta_w$ & - & 1.8 (0.9) & - & 12.6 (6.3) \\
        $\beta$ & 11 & 1 & 11 & 1 \\
        $\gamma$ & 5.5 & 0 & 5.5 & 0 \\
        $cc_{\mathrm{exp}}$ & 4151 $\pm$ 64& 2339 $\pm$ 48& 2798 $\pm$ 53& 2525 $\pm$ 50\\
        \hline
    \end{tabular}
    \caption{\textbf{Overview of the main parameters.} In this table, we report all the relevant hyperparameters used in the experimental implementation of the QON and QOSN models, and, in the last row, the average coincidence counts detected in the experimental implementations. As for the learning rates $\eta_\lambda$. $\eta_b$, and $\eta_w$, the numbers between parentheses report the corresponding updated values in the second half of the training procedure.}
\end{table}

\section*{Acknowledgments}

This work is supported by the ERC Advanced Grant QU-BOSS (QUantum advantage via non-linear BOSon Sampling, Grant No. 884676), by the project QU-DICE, Fare Ricerca in Italia, Ministero dell'istruzione e del merito, code: R20TRHTSPA, and by the PNRR MUR project PE0000023-NQSTI (Spoke 4).

\bibliographystyle{naturemag}
\bibliography{biblio}

\end{document}


\beginsupplement

\title{\textit{Supplementary Information for}:\\ Quantum Optical Neuron for Image Classification via Multiphoton Interference}

\author{Giorgio Minati}
\affiliation{Dipartimento di Fisica, Sapienza Universit\`{a} di Roma, Piazzale Aldo Moro 5, I-00185 Roma, Italy}

\author{Simone Roncallo}
\affiliation{Dipartimento di Fisica, Universit\`{a} degli Studi di Pavia, Via Agostino Bassi 6, I-27100 Pavia, Italy}

\author{Simone Scrofana}
\affiliation{Dipartimento di Fisica, Sapienza Universit\`{a} di Roma, Piazzale Aldo Moro 5, I-00185 Roma, Italy}

\author{Angela Rosy Morgillo}
\affiliation{Dipartimento di Fisica, Universit\`{a} degli Studi di Pavia, Via Agostino Bassi 6, I-27100 Pavia, Italy}

\author{Nicol\'{o} Spagnolo}
\affiliation{Dipartimento di Fisica, Sapienza Universit\`{a} di Roma, Piazzale Aldo Moro 5, I-00185 Roma, Italy}

\author{Chiara Macchiavello}
\affiliation{Dipartimento di Fisica, Universit\`{a} degli Studi di Pavia, Via Agostino Bassi 6, I-27100 Pavia, Italy}

\author{Lorenzo Maccone}
\affiliation{Dipartimento di Fisica, Universit\`{a} degli Studi di Pavia, Via Agostino Bassi 6, I-27100 Pavia, Italy}

\author{Valeria Cimini}
\email{valeria.cimini@uniroma1.it}
\affiliation{Dipartimento di Fisica, Sapienza Universit\`{a} di Roma, Piazzale Aldo Moro 5, I-00185 Roma, Italy}

\author{Fabio Sciarrino}
\affiliation{Dipartimento di Fisica, Sapienza Universit\`{a} di Roma, Piazzale Aldo Moro 5, I-00185 Roma, Italy}

\maketitle

\section{Details on the gradient descent optimization}

As described in the main text, the optimization of the model parameters is performed via gradient descent. In the overall gradient $\nabla H_\theta$ of the binary cross-entropy $H$, the only model-dependent contribution is given by:
\begin{equation}
\label{eq:model_dep_grad}
    \partial_\lambda z= 2 \mathrm{Re}{[\langle X , \mathcal{U}\rangle \langle X , \partial_\lambda \mathcal{U} \rangle]},
\end{equation}
where $\lambda$ denotes the model weights, $X$ and $\mathcal{U}$ represent the input and probe photons transfer functions, respectively, while $z$ is their overlap.\\
In order to have an explicit expression, an analytical model of the pixel-by-pixel modulation induced by the SLM can be introduced (further details in \cite{roncallo2025qon, roncallo2025shallow}) via a combination of top-hat functions. Defining pixel as $L \times L$ squares centered in position $r_{\mu \nu} = (\mu + 1/2, \nu + 1/2)L$, we can describe the weights spatial profile with the following expression:
\begin{equation}
\label{eq:top_hat}
    \mathcal{U}_\lambda (r) = \sum_{\mu \nu} u(r - r_{\mu \nu}) \frac{\lambda_{\mu \nu}}{\left \lVert \lambda \right \rVert},
\end{equation}
where $\left \lVert \lambda \right \rVert\equiv \sum_{\mu \nu} \lambda^2_{\mu \nu}$ and $u(r) \equiv \Theta(r + L/2) - \Theta(r - l/2)$, with $\Theta$ denoting the Heaviside step function. Moreover, we designed the SLM phase masks in such a way that the input photon carries the image information via amplitude encoding, while the spatial profile of the phase remains flat, for which reason we can approximate the spatial profile of the input photon to be real. Under this assumption, Eq.\eqref{eq:top_hat} allows us to rewrite Eq.\eqref{eq:model_dep_grad} as:

\begin{equation}
\label{eq:d_mu_nu}
    \frac{\partial z}{\partial \lambda_{\mu \nu}} \simeq 2 \frac{\sqrt{z}}{\left \lVert \lambda \right \rVert} \left[ (u \star X)(r_{\mu \nu}) - \sqrt{z} \frac{\lambda_{\mu \nu}}{\left \lVert \lambda \right \rVert} \right],
\end{equation}
in which $\star$ represents the cross-correlation operation.
Equation \eqref{eq:d_mu_nu} shows that the model-dependent part of the gradient requires, in principle, knowledge of both the input and probe spatial profiles, with the latter needing to be updated at each epoch as the weights change. In our experimental implementation, we circumvent the time cost and setup complexity associated with this continuous profile characterization by approximating the amplitude-encoded profiles using the corresponding "theoretical" discrete $32 \times 32$ image and weight arrays.

\begin{figure}
    \centering
    \includegraphics[width=0.99\linewidth]{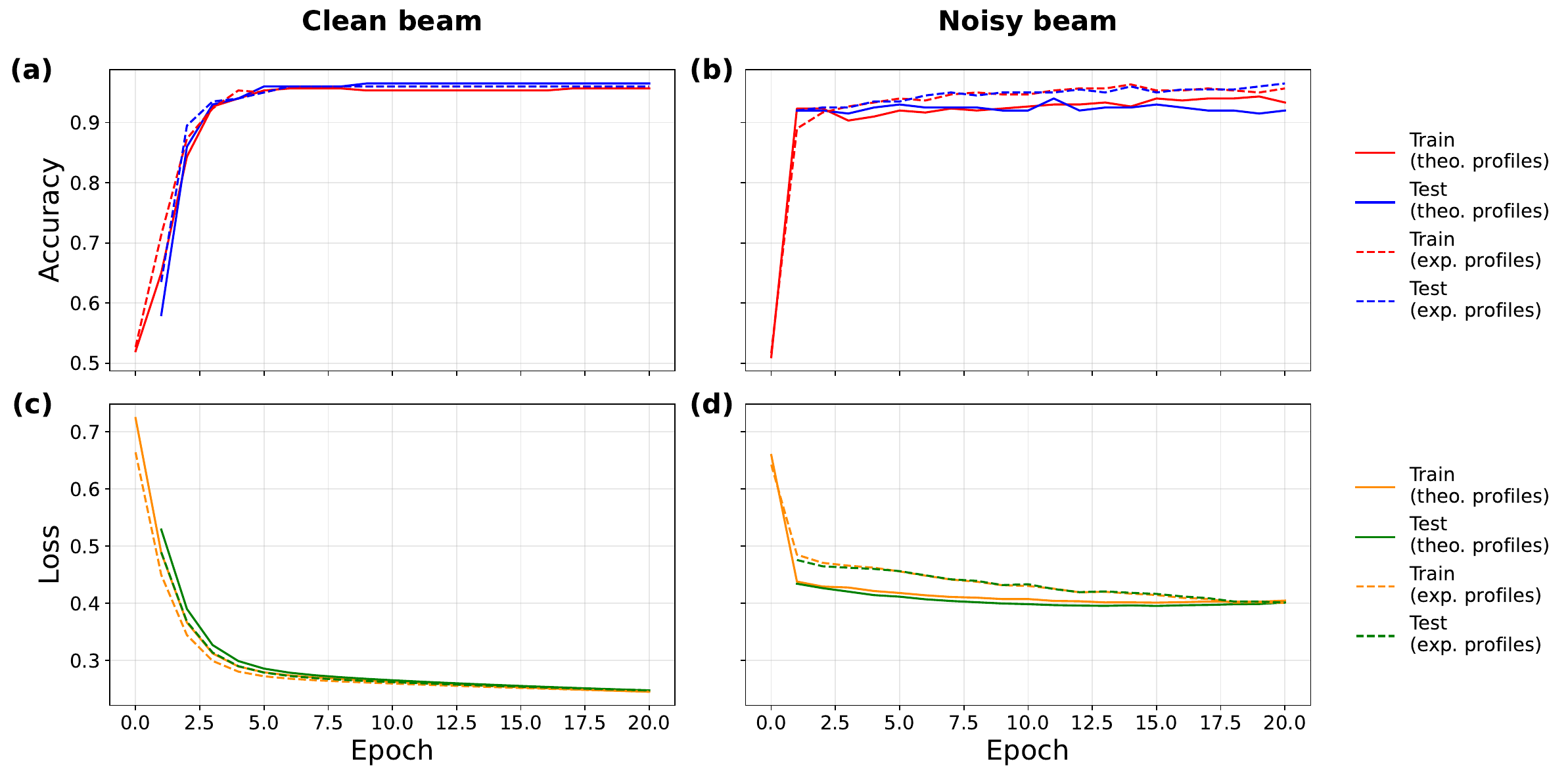}
    \caption{\textbf{QON performances with and without full profile characterization.} In panels \textbf{(a)}-\textbf{(b)} and \textbf{(c)}-\textbf{(d)}, we respectively report the accuracy and loss obtained when training the QON to classify ``sneaker'' and ``bag'' images from the fashion-MNIST dataset for 20 epochs, comparing the results achieved with noiseless (left panels) or noisy (right panels) spatial profiles. The train and test accuracies (losses) are reported as red (orange) and blue (green) lines. Instead, solid and dashed lines respectively correspond to the results obtained by performing the gradient descent optimization with the exact simulated input and probe profiles or the corresponding theoretical images and weights.}
    \label{fig:theo_vs_exp}
\end{figure}
In Fig.\ref{fig:theo_vs_exp}, we simulate the effects of this approximation under different noise conditions for the task of classifying ``sneaker'' and ``bag'' images randomly sampled from the Fashion-MNIST dataset. To remain consistent with our experimental conditions, we use 100 training and 40 test samples. We simulate the QON model by encoding the images onto photons with an initially gaussian spatial profile, using the technique described in \cite{wu2015simultaneous}, and numerically computing the final profiles after propagation from the SLM to the beam-splitter (BS) plane. As illustrated in Figs.\ref{fig:theo_vs_exp}a and \ref{fig:theo_vs_exp}b, when we consider noiseless profiles, the approximation has a negligible effect, since the absence of noise allows us to encode the image and weights onto the spatial profiles with minimal error. Conversely, in Figs.\ref{fig:theo_vs_exp}c and \ref{fig:theo_vs_exp}d, we consider beams subjected to local phase noise uniformly sampled from $\Delta \phi \in [0, \pi]$. In this scenario, the approximation does not account for the phase noise, effectively updating the weights using a gradient contribution in Eq.\eqref{eq:d_mu_nu} that differs slightly from one derived via full experimental profile characterization. Nonetheless, despite this non-negligible noise, the approximation does not significantly degrade the classification performance of the QON.

\section{QON performance under noisy conditions}
\subsection{Finite statistics effects}
\label{sec:finite_stat}
Under realistic experimental conditions, we are limited to finite detector integration times, meaning that the visibilities must be estimated from a finite number of coincidence counts. To quantify the impact of finite statistics on the QON classification performance, we simulate the training procedure by adding poissonian noise to the coincidence counts. Specifically, we use training and test sets consisting of 100 randomly chosen fashion-MNIST images (within the ``sneaker'' and ``bag'' classes) each, and train the model for 20 epochs.
\begin{figure}
    \centering
    \includegraphics[width=0.57\linewidth]{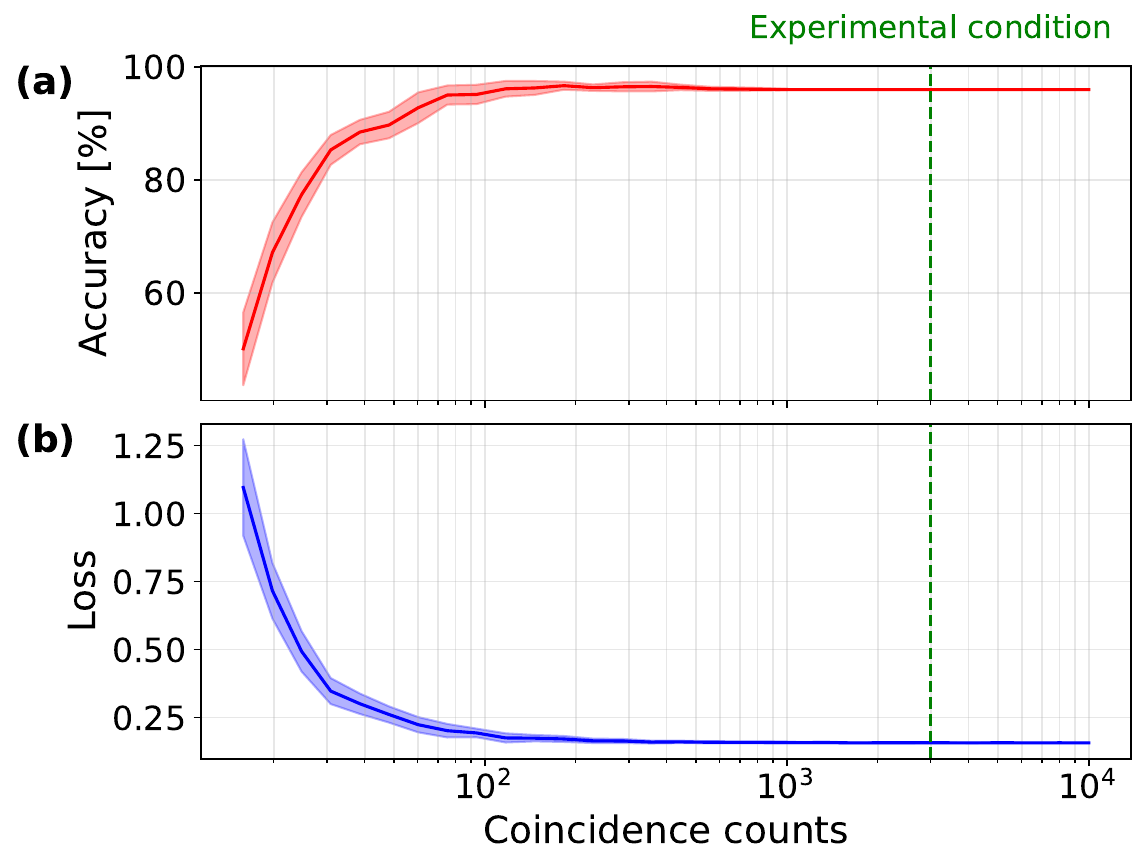}
    \caption{\textbf{Effects of finite statistics in the QON.} 
    Panels \textbf{(a)} and \textbf{(b)} show the accuracy and loss, respectively, as a function of the coincidence counts per acquisition for the QON model. In panel \textbf{(a)}, the red solid line represents the average test accuracy, with the shaded area indicating the standard deviation over 20 independent repetitions of a 20-epoch training session using a training set of 100 samples. Analogously, panel \textbf{(b)} reports the average and standard deviation of the loss function, depicted by the blue solid line and corresponding shaded area.
    }
    \label{fig:stat_finita}
\end{figure}
The average accuracy and loss over 20 repetitions of such noisy training are reported as red and blue solid lines in Fig.\ref{fig:stat_finita}a-b, respectively, while the shaded areas depict the corresponding standard deviation. This result shows that the statistical noise becomes irrelevant whenever the visibilities are estimated with at least a few hundred coincidence counts. In our experiment, we work with an average signal of $\sim 3000$, defined as the measured coincidence counts without HOM interference, averaged over the entire dataset, per visibility estimation (the average signals of the individual measurements are reported in the Materials and Methods section of the main text). In Fig.\ref{fig:stat_finita}, we can see how this quantity is well above the regime in which the QON performance converges.
\begin{figure}
    \centering
    \includegraphics[width=0.82\linewidth]{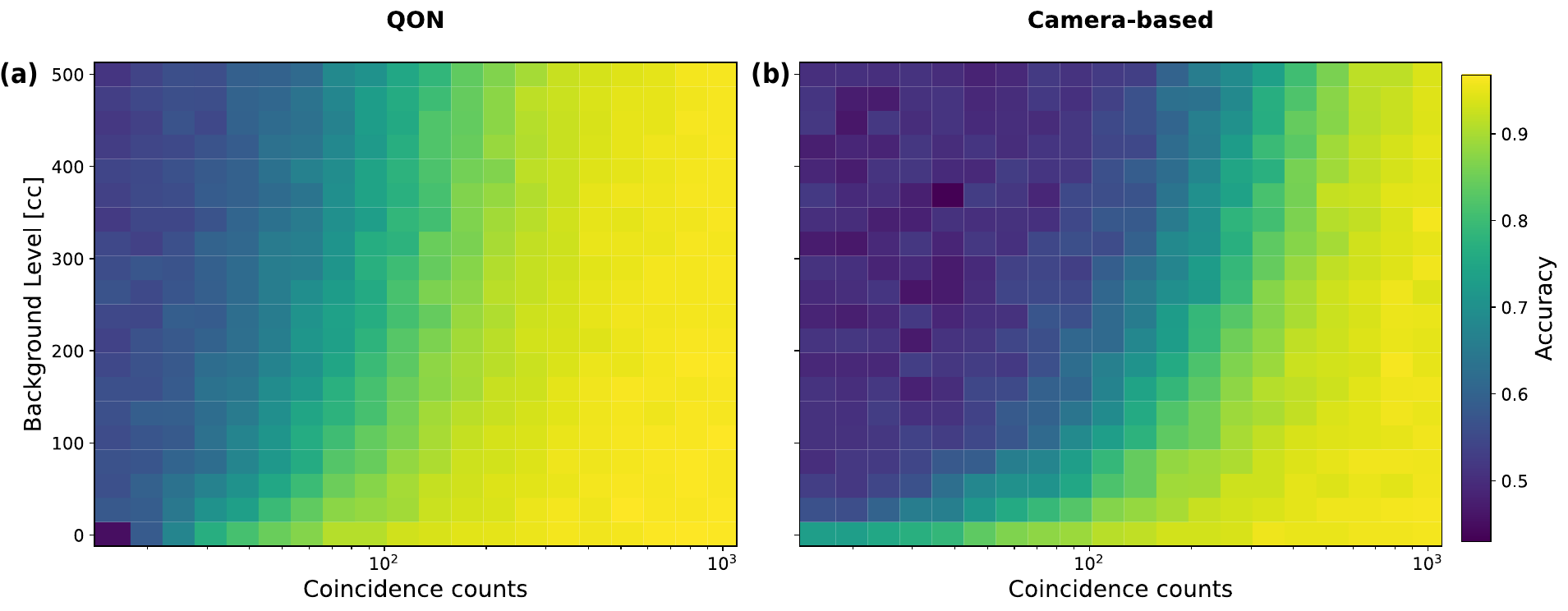}
    \caption{\textbf{Classification performance of the QON and camera-based models under statistical and background noise.} In panels \textbf{(a)} and \textbf{(b)}, we report the test classification accuracies achieved with a finite number of coincidence counts and constant background noise for the QON model and the camera-based approach, respectively. While in the former the classification is based on global visibility measurements, in the latter, the detection has a spatial resolution of $32 \times 32$ pixels. As a benchmark, we use training and test datasets of 100 images sampled from the Fashion-MNIST ``sneaker'' and ``bag'' classes.} 
    \label{fig:QON_vs_camera}
\end{figure}
Furthermore, we can evaluate the QON model's resource efficiency by comparing it to a camera-based model where images are reconstructed using spatially resolved detection on a $32 \times 32$ grid, rather than measuring coincidence counts with bucket detectors. The reconstructed profile is then classified using a single perceptron with the same number of weights and sigmoid activation function as the QON model. To make the noise model more realistic, we also introduce constant background noise, uniformly distributed across the camera-based model's detection pixels. As a benchmark, we classify ``sneaker'' and ``bag'' images from the Fashion-MNIST dataset, using 100 randomly chosen samples for both the training and test sets. In Fig.\ref{fig:QON_vs_camera}, we illustrate the results of this comparison, showing that the QON model is more resilient than the camera-based model to background noise and low detection statistics, while both achieve a similar classification accuracy in the limit of high statistics. This result is expected since, under the same conditions, the coincidence counts are spread across all the pixels in the camera-based model, thereby increasing the individual pixel statistical noise.

\subsection{Limited visibility}
As discussed in Sec.~\ref{sec:finite_stat}, while a sufficiently high coincidence count rate can significantly reduce statistical noise, another crucial source of degradation arises from the non-ideal overlap of the signal and idler quantum states. This mismatch originates from experimental imperfections in the calibration of polarization, spectral profiles, relative delay, and spatial alignment, all of which directly limit the maximum achievable HOM visibility. To test the resilience of the QON under these conditions, we trained the model to classify Fashion-MNIST images while artificially rescaling all measured visibilities by a constant factor, $\eta_{\mathrm{vis}}$. For this simulation, we use 200 randomly chosen fashion-MNIST samples from the ``sneaker'' and ``bag'' classes, 100 used as the training set and 100 as the test set.  
\begin{figure}
    \centering
    \includegraphics[width=0.9\linewidth]{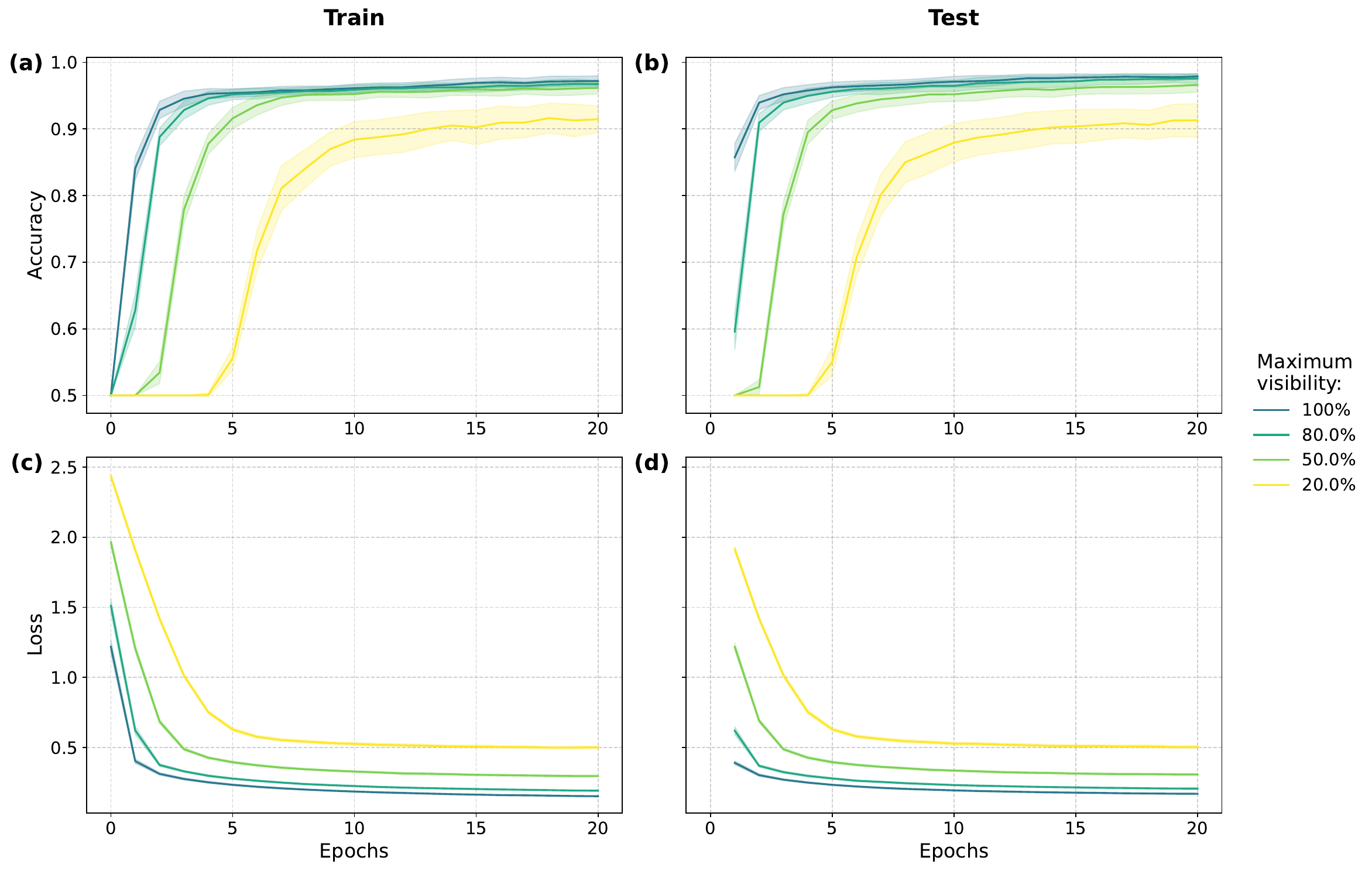}
    \caption{\textbf{Effects of limited visibility on QON performance.} In panels \textbf{(a)} and \textbf{(b)}, we respectively report the train and test QON accuracies in classifying ``sneaker'' and ``bag'' images from fashion-MNIST, while reducing all the visibilities by a limiting factor $\eta_{\mathrm{vis}}$. We report, using different colors, different instances of visibility reduction, namely $\eta_{\mathrm{vis}}=$1, 0.8, 0.5, and 0.2, illustrating as solid lines and shaded areas the averages and standard deviations over 100 independent repetitions of the simulation with a statistical poissonian noise on on an average coincidence count per sample of $cc=3000$. Analogously, in panels \textbf{(c)} and \textbf{(d)}, we report the corresponding values of the loss function on the training and test set, respectively.}
    \label{fig:limited_vis}
\end{figure}
In Fig.~\ref{fig:limited_vis}, we present results for varying levels of visibility degradation ($\eta_{\mathrm{vis}} = 1, 0.8, 0.5$, and $0.2$), combined with the statistical noise generated by an average coincidence counts per sample of $cc = 3000$, which closely matches our experimental conditions. Remarkably, the QON model demonstrates strong resilience to limited visibility, with classification performance remaining almost unaffected for $\eta_{\mathrm{vis}} \geq 0.5$, and even under extreme overlap reduction ($\eta_{\mathrm{vis}} = 0.2$), the model still correctly classifies more than 90\% of the samples. 
\clearpage

\bibliographystyle{naturemag}
\bibliography{biblio}